\documentclass[runningheads]{llncs}
\usepackage{graphicx}
\usepackage{amsmath}

\def\lpmln{{\rm LP}^{\rm{MLN}}}
\def\mo#1{{\mathsf{#1}}}
\def\beq{\begin{equation}}
\def\eeq#1{\label{#1}\end{equation}}
\def\o{\overline}
\def\ba{\begin{array}}
\def\ea{\end{array}}
\def\ar{\leftarrow}

\def\o{\overline}
\def\ar{\leftarrow}
\def\bi{\begin{itemize}}
\def\ei{\end{itemize}}
\def\beq{\begin{equation}}
\def\eeq#1{\label{#1}\end{equation}}
\def\ba{\begin{array}}
\def\ea{\end{array}}
\def\i#1{\hbox{\it #1\/}}

\def\sm{\rm SM}

\def\no{\i{not}}

\def\ar{\leftarrow}
\def\rar{\rightarrow}
\def\lrar{\leftrightarrow}
\def\no{\i{not}}

\def\i#1{\hbox{\itshape #1\/}}

\DeclareSymbolFont{AMSa}{U}{msa}{m}{n}
\DeclareMathSymbol{\square}{\mathord}{AMSa}{"03}

\def\lpmln{{\rm LP}^{\rm{MLN}}}

\newtheorem{thm}{Theorem}

\begin{document}
\title{Strong Equivalence for $\lpmln$ Programs}

\author{Man Luo}
\authorrunning{}

\institute{Arizona State University, Tempe AZ 85281, USA \\
\email{mluo26@asu.edu}\\
}
\maketitle              
\begin{abstract}
Strong equivalence is a well-studied and important concept in answer set programming (ASP). $\lpmln$ is a probabilistic extension of answer set programs with the weight scheme adapted from Markov Logic. Because of the semantic differences, strong equivalence for ASP does not simply carry over to $\lpmln$. I study the concept of strong equivalence in $\lpmln$ with the goal of extending strong equivalence to $\lpmln$ programs. My study shows that the verification of strong equivalence in $\lpmln$ can be reduced to equivalence checking in classical logic plus weight consideration.The result allows us to leverage an answer set solver for checking strong equivalence in $\lpmln$. Furthermore, this study also suggests us a few reformulations of the $\lpmln$ semantics using choice rules, logic of here and there, and the second-order logic. I will present my work result of strong equivalence for $\lpmln$ and talk about my next steps for research: one is approximately strong equivalence, and another is the integration of fuzzy logic with neural network.  

\end{abstract}
\section{Introduction}
$\lpmln$ is a probabilistic extension of answer set programs with the weight scheme adapted from Markov Logic \cite{richardson06markov}.  An $\lpmln$ program defines the probability distribution over all ``soft'' stable models, which do not necessarily satisfy all rules in the program, but the more rules with the bigger weights they satisfy, the bigger their probabilities.

$\lpmln$ turns out to be highly expressive to embed several other probabilistic logic languages, as more results are built upon $\lpmln$, it becomes more important to consider the equivalence between different $\lpmln$ programs. As with answer set programs,  $\lpmln$ programs $\mo{F}$ and $\mo{G}$ that have the same soft stable models with the same probability distribution are not necessarily equivalent in a stronger sense. When we add the same program $\mo{H}$ to each of $\mo{F}$ and $\mo{G}$, the resulting programs may have different soft stable models and different probability distributions. 

However, because of the semantic differences, strong equivalence for answer set programs does not simply carry over to $\lpmln$. Based on this observation, I study the notion of strong equivalence for $\lpmln$ and characterize this concept. 

The paper is organized as follows: Section~\ref{sec:se} presents the background of my research. Section~\ref{sec:soft-ht} explains the central question of my research. Section \ref{sec:se-solver} shows the main accomplishment of my current work and the plan for the future work.

\section{Background: Language $\lpmln$}\label{sec:se}

 An $\lpmln$ program is a finite set of weighted formulas  $w: R$ where $R$ is a propositional formula 
 and $w$ is a real number (in which case, the weighted rule is called {\em soft}) or $\alpha$ for denoting the infinite weight (in which case, the weighted rule is called {\em hard}). 
 
For any $\lpmln$ program $\mo{F}$ and any set $X$ of atoms,  
$\o{\mo{F}}$ denotes the set of usual (unweighted) formulas obtained from $\mo{F}$ by dropping the weights, and
$\mo{F}_X$ denotes the set of $w: R$ in $\mo{F}$ such that $X\models R$.

Given an $\lpmln$ program $\mo{F}$, $\sm[\mo{F}]$ denotes the set of {\em soft stable models}: 
\[
\ba l
\{X\mid \text{$X$ is a (standard) stable model of $\o{\mo{F}_X}$}\}.
\ea
\]

By $\i{TW}(\mo{F})$ (``Total Weight" of $\mo{F}$) we denote the expression $exp({\sum\limits_{  w:R\in \mo{F}} w})$.
For any interpretation $X$, the weight of an interpretation $X$, denoted $W_{\mo{F}}(X)$, is defined as
{
\[
 W_{\mo{F}}(X) =
\begin{cases}
    \i{TW}(\mo{F_X})  & 
      \text{if $X\in\sm[\mo{F}]$}; \\
  0 & \text{otherwise},
\end{cases}
\]
}
and the probability of $X$, denoted $P_\mo{F}(X)$, is defined as
{\small
\[
  P_\mo{F}(X)  = 
  \lim\limits_{\alpha\to\infty} \frac{W_\mo{F}(X)}{\sum\limits_{Y\in {\rm SM}[\mo{F}]}{W_\mo{F}(Y)}}. 
\]
}

\section{Central Question: Strong Equivalence for $\lpmln$}\label{sec:soft-ht}

Strong equivalence is an important concept in the theory of answer set programming. Informally speaking, logic programs $\mo{P}$ and $\mo{Q}$ are strongly equivalent if, given any program $\mo{R}$, programs $\mo{P} \cup \mo{R}$ and $\mo{Q} \cup  \mo{R}$ have the same stable models. On the other hand, Logic of Here and There (logic $HT$) is proven to be useful for a monotonic basis for checking strong equivalence \cite{lif01}, and equilibrium models \cite{pearce06equilibrium} are defined as a special class of minimal models in logic $HT$. 

However, because of the semantic differences, strong equivalence for answer set programs does not simply carry over to $\lpmln$.  First, weights play a role. Even for the same structure of rules, different assignments of weights make the programs no longer strongly equivalent. Also, due to the fact that soft stable models do not have to satisfy all rules, strongly equivalent answer set programs do not simply translate to strongly equivalent $\lpmln$ programs. For instance, 
$\{a\lor b, \ \ \bot\ar a, b\}$ is strongly equivalent to 
$\{a\ar \no\ b, \ \ b \ar \no\ a, \ \ \bot\ar a,b \}$, but its $\lpmln$ counterpart
$\{\alpha: a\lor b, \ \  \alpha:\bot\ar a, b\}$ is not strongly equivalent to 
$\{\alpha: a\ar \no\ b, \ \ \alpha: b \ar \no\ a, \ \ \alpha:\bot\ar a,b \}$: if we add 
$\{\alpha: a\ar b,\ \  \alpha: b\ar a\}$ to each of them, $\{a,b\}$ is a soft stable model of the former (by disregarding the rule $\alpha: \bot\ar a,b$) but not of the latter

I extend the notion of strong equivalence to $\lpmln$, and show that the verification of strong equivalence in $\lpmln$ can be reduced to equivalence checking in classical logic plus weight consideration. We also extend the logic of here and there to weighted rules, which provides a monotonic basis of checking strong equivalence. 
The study of strong equivalence suggests us a few reformulations of the $\lpmln$ semantics using choice rules, logic of here and there, and second-order logic, which present us useful insights into the semantics. 

Definition 1 is the notion of strong equivalence in $\lpmln$ in terms of probability distribution. 

\begin{definition} \label{def:se}
$\lpmln$ programs $\mo{F}$ and $\mo{G}$ are called {\em strongly equivalent} to each other if, for any $\lpmln$ program $\mo{H}$,
\[
   P_{\mo{F}\cup \mo{H}}(X) = P_{\mo{G}\cup \mo{H}}(X)
\]
for all interpretations $X$.
\end{definition}
The following theorem shows a characterization of strong equivalence that does not need to consider adding all possible $\lpmln$ program $\mo{H}$, which can be reduced to equivalence checking in classical logic plus weight checking.

For any $\lpmln$ program $\mo{F}$ and any set $X$ of atoms,  
$\o{\mo{F}}$ denotes the set of usual (unweighted) formulas obtained from $\mo{F}$ by dropping the weights, and
$\mo{F}_X$ denotes the set of $w: R$ in $\mo{F}$ such that $X\models R$.

\begin{thm} \label{thm:se-char} 
For any $\lpmln$ programs $\mo{F}$ and $\mo{G}$, program $\mo{F}$ is strongly equivalent to $\mo{G}$ if and only if there is a w-expression $c$ such that for every interpretation $X$, 
\begin{enumerate}
\item $TW(\mo{F}_X) = c \times TW(\mo{G}_X)$, and 
\item $(\o{\mo{F}_X})^X$ and  $(\o{\mo{G}_X})^X$ are classically equivalent. 
\end{enumerate}
where $w$-expression is in the form of $e^{c_1+c_2\alpha}$, and $c_1$ is a real number counting the weight of soft rules, $c_2$ is an integer counting the weight of hard rules. 
\end{thm}

\begin{example}\label{ex:main}
Consider two programs 
\[
\ba {lrcllrcl}
\mo{F} ~~ & 0: & \neg a  & \hspace{5mm} & \mo{G} ~~  & 2: & \neg a \lor b \\
            & 2: & b\ar a  &                         &              &1: &  a\lor \neg a  \\
            & 3: & a \ar \neg \neg a \\
\ea 
\] 
Programs  $\mo{F}$ and $\mo{G}$ are strongly equivalent to each other. The following table shows $\mo F$ and $\mo G$ statisfy two conditions in Theorem~\ref{thm:se-char}.

\begin{table}[h!] 
\begin {center}
\begin{tabular}{c|c|c|c|c}

$X$  & $TW(\mo{F}_X)$ & $TW(\mo{G}_X)$ & $(\o{\mo{F}_X})^X$ & $(\o{\mo{G}_X})^X$ \\ \hline \hline
$\phi$ &$e^{5}$ & $e^{3}$ & $\top$ & $\top$ \\
\hline
$\{a\}$ &$e^3$ & $e^1$ & $a$ & $a$ \\
\hline
$\{b\}$ & $e^5$ & $e^3$ & $\top$ & $\top$ \\
\hline
$\{a, b\}$& $e^5$ & $e^3$  & $a\land b$ & $a\land b$
\end{tabular}
\end{center}
\vspace{-3mm}
\caption{$(\o{\mo{F}_X})^{X}$ and $(\o{\mo{G}_X})^{X}$}
\vspace{-2mm}
\label{tab:table1}
\end{table}

From the first and the second column, it is easy to see that $TW({\mo{F}_X}) = e^2 \times TW(\mo{G}_X)$, so the first condition in Theorem 1 is satisfied. The third and forth column show that the second condition in Theorem 1 is satisfied. However, if we replace rule \ \ $3:\ \ a \ar \neg\neg a$\ \  in $\mo{F}$ with \ \ $3:\ \ a \ar a$\ \  to result in $\mo{F}'$, then $\mo{F}'$ and $\mo{G}$ are not strongly equivalent: for 
$$
\mo{H}=\{1 :a \leftarrow b, \ \ 1: b \leftarrow a\}
$$ 
$\{a, b\}$ is a soft stable model for $\mo{G}\cup\mo{H}$ with the weight $e^5$, but it is not a soft stable model for $\mo{F'}\cup\mo{H}$, so its weight is 0. 
In accordance with Theorem~\ref{thm:se-char}, 
$(\o{\mo{F'}_{\{a, b\}}})^{\{a, b\}}$ is
not equivalent to 
$(\o{\mo{G}_{\{a, b\}}})^{\{a, b\}}$. 
The former is equivalent to $\{  b \leftarrow a\}$, 
and the latter is equivalent to 
$\{a \land b\}$.

Even if the programs have the same soft stable models, the different weight assignments may make them not strongly equivalent. For instance, replacing the first rule in $\mo{G}$ by \hbox{$3: \neg a \lor b$} to result in $\mo{G}'$, we have $TW(\mo{F}_{\phi}) = e^1 \times TW(\mo{G'}_{\phi})$ and $TW(\mo{F}_{\{a\}}) = e^2 \times TW(\mo{G'}_{\{a\}})$, so there is no $w$-expression $c$ such that $TW(\mo{F}_X) = c \times TW(\mo{G'}_X)$.

\end{example}
Based on the concepts of choice rules,  logic HT\cite{hey30} and the second order, we present the theorem on soft stable, in which every item is equivalent to each other. Before  introducing the theorem, we define the definition of choice formula and $\Delta_{\bf P'}(F)$.  For any propositional formula $F$, by $\{F\}^{\rm ch}$ we denote the choice formula $F\lor\neg F$. 

Let ${\bf p}$ be the propositional signature.
Let ${\bf p}'$ be the set of atoms $p'$ where $p\in {\bf p}$.
For any formula $F$, $\Delta_{{\bf p}'}(F)$ is defined recursively:
\begin{itemize}
\item $\Delta_{{\bf p}'}(p) = p'$ for any atomic formula $p\in {\bf p}$;
\item $\Delta_{{\bf p}'}(\neg F) = \neg F$;
\item $\Delta_{{\bf p}'}(F \land G) = \Delta_{{\bf p}'}(F) \land \Delta_{{\bf p}'}(G)$;
\item $\Delta_{{\bf p}'}(F \lor G) = \Delta_{{\bf p}'}(F) \lor \Delta_{{\bf p}'}(G)$;
\item $\Delta_{{\bf p}'}(F \rightarrow G) = (\Delta_{{\bf p}'}(F) \rightarrow \Delta_{{\bf p}'}(G)) \land (F \rightarrow G)$.
\end{itemize}

\medskip\noindent
{\bf Theorem on Soft Stable Models}\ \ \ 

For any $\lpmln$ program $\mo{F}$ and $\mo{G}$,
the following conditions are equivalent. By $\mo{F}^X$, we denote the reduct of $\mo{F}$ obtained from $\mo{F}$ by replacing every maximal subformula of $\mo{F}$ that is not satisfied by $X$ with $\bot$.
\begin{enumerate}
\item[(a)]  For any $\lpmln$ program $\mo{H}$, programs $\mo{F}\cup\mo{H}$ and $\mo{G}\cup\mo{H}$ have the same soft stable models. 

\item[(b)] For any set $X$ of atoms, $(\o{\mo{F}_X})^X$ and $(\o{\mo{G}_X})^X$ are classically equivalent.

\item[(c)] For any set $X$ of atoms, $(\{\o{\mo{F}}\}^{\rm ch})^X$ and $(\{\o{\mo{G}}\}^{\rm ch})^X$ are classically equivalent.

\item[(d)] $\mo{F}$ and $\mo{G}$ have the same set of soft $HT$ models.

\item[(e)]  $(\{\o{\mo{F}}\}^{\rm ch}) \lrar (\{\o{\mo{G}}\}^{\rm ch})$ is provable in $HT$. 

\item[(f)]  For any set $X$ of atoms, $\{p'\rar p \mid p \in {\bf p}\}$ 
entails $\Delta(\o{\mo{F}_X})\lrar \Delta(\o{\mo{G}_X})$
 (in the sense of classical logic).

\item[(g)]
$\{p'\rar p \mid p \in {\bf p}\}$ entails
   $ \Delta(\{\o{\mo{F}}\}^{\rm ch})\lrar\Delta(\{\o{\mo{G}}\}^{\rm ch})$
(in the sense of classical logic).

\end{enumerate}

\section{Accomplishment and Future work} \label{sec:se-solver}

\subsection{Accomplishment}

Theorem 1 shows a characterization of strong equivalence that does not need to consider adding all possible $\lpmln$ programs $\mo{H}$.  Similar  to Proposition~2 from \cite{fer05}, it shows that the verification of strong equivalence in $\lpmln$ can be reduced to equivalence checking in classical logic plus weight checking.

I am still at the beginning stage in research. I get familiar with $\lpmln$ language  by the study of strong equivalence and it indeed gives me more insight and I feel quite interested in the area of knowledge representation and reasoning. In the following, I will show two interesting topics that I will investigate in my next step.
\subsection{Future work}

\subsubsection{Approximate strong equivalent.}

 We plan to extend the work to approximate strong equivalence, where the probability distributions may not necessarily be identical but allowed to be slightly different with some error bound. Approximate strong equivalent for $\lpmln$ will have more flexibility. 
One application of approximate strong equivalence is weight learning in $\lpmln$. More specifically, due to the difference of initialized value of weights and the noise existing in the data, there is no guarantee that the weight learned from the data will be exactly the same. In such case, we should allow some certain bound so that the different set of weights can be approximate strong equivalence.

\subsubsection{Integration of symbolic and sub-symbolic.}

In symbolic systems, knowledge is encoded in terms of explicit structure(rules) and inferences are  based on this structure(rules). Neural networks provide a powerful mechanism for learning patterns from massive data. Although neural network can learn the model from data, it has difficulty with high level reasoning. The integration of these two systems can take advantage from both sides: symbolic system can do the reasoning based on model, and neural networks can learn the model from the data. 
One obvious advantage of this integration is that the rules defined in symbolic system can "guide" neural network learn the relation among data.
Recently, more and more studies focus on this field, such as Neural tensor network(LTN)\cite{donadello2017logic}, TensorLog\cite{cohen2016tensorlog} etc. For instance, LTN is an interesting framework, in which it grounds the terms with vectors representing the features of the objects and grounds the clauses with real value in the interval [0,1] representing the truth degree of the clauses. By maximizing the truth degree of the clauses, LTN encodes the knowledge into neural networks. I am applying LTN to different tasks to get some insight with the goal of inventing an integration of symbolic and subsymbolic that works well for different domains.


\begin{thebibliography}{1}
\providecommand{\url}[1]{\texttt{#1}}
\providecommand{\urlprefix}{URL }
\providecommand{\doi}[1]{https://doi.org/#1}

\bibitem{cohen2016tensorlog}
Cohen, W.W.: Tensorlog: A differentiable deductive database. arXiv preprint
  arXiv:1605.06523  (2016)

\bibitem{donadello2017logic}
Donadello, I., Serafini, L., Garcez, A.D.: Logic tensor networks for semantic
  image interpretation. arXiv preprint arXiv:1705.08968  (2017)

\bibitem{fer05}
Ferraris, P.: Answer sets for propositional theories. In: Proceedings of
  International Conference on Logic Programming and Nonmonotonic Reasoning
  ({LPNMR}). pp. 119--131 (2005)

\bibitem{hey30}
Heyting, A.: Die formalen {R}egeln der intuitionistischen {L}ogik.
  Sitzungsberichte der Preussischen Akademie von Wissenschaften.
  Physikalisch-mathematische Klasse pp. 42--56 (1930)

\bibitem{lif01}
Lifschitz, V., Pearce, D., Valverde, A.: Strongly equivalent logic programs.
  ACM Transactions on Computational Logic  \textbf{2},  526--541 (2001)

\bibitem{pearce06equilibrium}
Pearce, D.: Equilibrium logic. Annals of Mathematics and Artificial
  Intelligence  \textbf{47}(1-2),  3--41 (2006)

\bibitem{richardson06markov}
Richardson, M., Domingos, P.: Markov logic networks. Machine Learning
  \textbf{62}(1-2),  107--136 (2006)

\end{thebibliography}

\end{document}